\documentclass[usenatbib]{mn2e}
\usepackage{natbib}
\usepackage{textcomp}
\usepackage[dvips]{graphicx}
\usepackage{aas_macros}
\usepackage{amsmath}
\usepackage{amssymb}
\bibpunct[, ]{(}{)}{;}{a}{}{,}

\newcommand{\pct}{\,\%}
\newcommand{\grad}{\mbox{\textdegree}}
\newcommand{\dez}[1]{{}\cdot10^{#1}\,}
\newcommand{\pma}[2]{{}^{+#1}_{-#2}}
\newcommand{\unit}[1]{\mbox{$\,\mathrm{#1}$}}
\newcommand{\ecc}{e} 

\newcommand{\mearth}{M_{\oplus}}
\newcommand{\msun}{M_{\sun}}
\newcommand{\mjup}{M_\mathrm{J}}
\newcommand{\tmsun}{\mbox{$\msun$}}
\newcommand{\tmjup}{\mbox{$\mjup$}}
\newcommand{\mdisk}{M_\mathrm{d}}

\newcommand{\psig}{q_{\Sigma}}
\newcommand{\pt}{q_{T}}
\begin{document}
\title[Formation of hot misaligned planets]{A natural formation scenario for misaligned and short-period eccentric extrasolar planets}
\author[Thies et al.]{I. Thies$^{1}$\thanks{E-mail: ithies@astro.uni-bonn.de},
P. Kroupa$^{1}$\thanks{pavel@astro.uni-bonn.de},
S.~P. Goodwin$^{2}$,
D. Stamatellos$^{3}$,
A.~P. Whitworth$^{3}$\\
$^{1}$Argelander-Institut f\"ur Astronomie (Sternwarte), Universit\"at Bonn, Auf dem H\"ugel 71, D-53121 Bonn, Germany\\
$^{2}$Department of Physics and Astronomy, University of Sheffield, Sheffield S3 7RH, UK\\
$^{3}$School of Physics \& Astronomy, Cardiff University, Cardiff CF24 3AA, UK}
\pagerange{1817--1822}
\volume{417}
\pubyear{2011}
\maketitle
\begin{abstract}
Recent discoveries of strongly misaligned transiting
exoplanets pose a challenge to the established planet formation theory
which assumes planetary systems to form and evolve in
isolation. However, the fact that the majority of stars actually do
form in star clusters raises the question how isolated forming
planetary systems really are. Besides radiative and tidal forces the
presence of dense gas aggregates in star-forming regions are potential
sources for perturbations to protoplanetary discs or systems. Here we
show that subsequent capture of gas from large extended accretion
envelopes onto a passing star with a typical circumstellar disc can
tilt the disc plane to retrograde orientation, naturally explaining
the formation of strongly inclined planetary systems. Furthermore, the
inner disc regions may become denser, and thus more prone to speedy
coagulation and planet formation.
Pre-existing planetary systems are compressed by gas inflows
leading to a natural occurrence of close-in misaligned hot Jupiters
and short-period eccentric planets.
The likelihood of such events mainly
depends on the gas content of the cluster and is thus expected to be
highest in the youngest star clusters.
\end{abstract}

\section{Introduction}\label{sec:intro}
The discoveries of retrograde or strongly misaligned
transiting exoplanets like WASP-17b and others
\citep{Heetal08,Gillon09,Joetal09,Naritaetal09,Pontetal09,Pontetal10,Winetal09a,Winetal09b,wasp17b_01,Triaudetal2010}
pose a challenge for established planet formation theories.
These classically assume the birth of planetary systems out
of collapsing protostellar cloud fragments. While contracting from
sub-parsec to milliparsec (or hundreds of AU) scale the conservation
of its initial angular momentum causes the cloud to spin and flatten,
while part of its mass settles towards the center, eventually igniting
the hydrogen burning in the stellar core.
Given such a protostar with a circumstellar
disc that contains a certain amount of dust, solid particles collide and
stick together, forming larger particles. This coagulation leads to bodies
(cores) of sufficient mass to accrete surrounding material (dust and even gas)
by gravity until no more disc material is available. This mechanism is probably
the dominant process of the formation of the Solar System, including the
Earth \citep{2008ASPC..398..235M}.
As an alternative, gravitational instability as a forming mechanism for giant
planets and brown dwarfs is being discussed \citep{2004ApJ...610..456B}, and
at least for brown dwarfs it has actually been shown to work
\citep{StHuWi07,Thiesetal2010}. More recent variations
of these models assume gravitational instability of the solid phase only
or dust trapping in vortices as a speed-up for coagulation.

Being isolated from any
external perturbation, everything inside this protostar-disc system
spins in the same direction, thus the forming planets orbit their host
star the same way. This simple model has been severely
questioned by the findings mentioned above.
The transiting exoplanet WASP-17b \citep{wasp17b_01},
for example, has been found to have a sky-projected inclination of
the orbital plane normal against the stellar spin axis of
$148.5\pma{5.1}{4.2}$ degrees. A number of other mis-aligned transiting
exoplanets have been discovered
\citep{Heetal08,Gillon09,Joetal09,Naritaetal09,Pontetal09,Pontetal10,Winetal09a,Winetal09b,wasp17b_01}.
These discoveries suggest that
misalignment between planetary orbits and the spin of their host star
(hereafter called spin-orbit misalignment) are quite common, at least among
close-in transiting planets for which the spin-orbit alignment can be measured
by using the Rossiter-McLaughlin Effect \citep{Rossiter1924,McLaughlin1924}.
The nearby $\upsilon$~And planetary system exhibits even mutually
inclined planetary orbits \citep{2010ApJ...715.1203M}.

Such planetary orbits cannot be explained by the standard planet
formation model according to which the conservation of angular momentum
forces any coplanar system to remain coplanar forever. There have been recent
suggestions that may shed some light into the possible mechanism.
\cite{FabTre07,Naozetal2011} deduce a formation scenario for misaligned ``hot jupiters''
through long-term mutual perturbations of two
planets (or an inner planet and an outer brown dwarf; for simplicity,
we use the term ``planet'' for all substellar companions in this
paper) orbiting the same star. Given an initial mutual inclination
between 40\grad\ and 140\grad\ the Kozai-Lidov mechanism leads to
secular oscillations of eccentricity vs. inclination. Tidal friction
circularises the orbit of the inner planet if its periastron
falls below a few stellar radii, eventually leaving the planet on a
close-in orbit with near-random inclination with respect to the
stellar spin. This scenario, however, requires a sufficient initial
mutual inclination of both planets.

Another possible mechanism is tilting the spin axis of the star.
Direct stellar encounters within a stellar radius (about 0.01~AU) can be
effectively ruled out, because they are improbable, even within stellar
clusters \citep{TKT05, TK07} and would destroy the disc.
Secular transfer of angular momentum between the protostellar spin and
the protoplanetary disc via magnetic fields have been investigated by
\citet{Laietal11}.

Multi-stage or episodic accretion of circumstellar material may
provide another viable mechanism for misaligned planetary systems.
The underlying idea is that
accretion from different sources (i.e. gas filaments, accretion envelopes)
from different directions may sever the classically assumed spin-orbit correlation.
The star itself may get its angular momentum from a different accretion event
than the bulk of circumstellar material.
\citet{BLP10} have analysed turbulent accretion events using data from
\citet{BBB03}. They deduce a high frequency of misaligned planets
as a consequence of multi-directional accretion. However, due to the limited
simulation data available, they were forced to make simplified assumptions
to what happens within the close vicinity of a star.
Our work aims to take a deeper insight into such scenarios by self-consistent
calculations of close interactions of a circumstellar disc with encountered
new material, and by also treating the effects on pre-existing planets.

As \citet{PTS08,OPE10} have shown, encounters between stars and pre-/protostellar
objects do occur frequently in dense stellar environments and may lead to
significant accretion bursts. Although the precise
mechanisms proposed \citep{PTS08} differ from ours, their calculations
emphasise the importance of encounter events to accretion processes.

In this paper the effects of gas capture onto a pre-existing circumstellar
disc are studied.
In Section \ref{sec:model} we depict the physical scenario while
Section \ref{sec:methods} briefly describes the numerical
methods used in our computations. The results are presented in Section
\ref{sec:results}.

\section{Model}\label{sec:model}
In our model, both the initial mutual
inclination as well as the inclination with respect to the stellar equator can be
naturally explained by multi-stage accretion during the formation
process of the planetary system. The scenario can be described as
follows: First, a star with a protoplanetary disc (PPD) is forming the
classical way, i.e. out of a collapsing spinning cloud fragment. Soon
after this the protostar passes another region in the star cluster
that contains dense gas, e.g. the protostellar cloud fragment or
accretion envelope of another star, hereafter called the target.
Accordingly, the star which encounters the target and its disc
are referred to as the bullet star and the bullet disc, respectively.
In our calculations, a massive circumstellar
disc around another star is used as the target.
While grazing the outer regions of
this target the star captures material from it at an angle that depends
on the encounter orbit and the arbitrarily chosen orientation of the target
and the bullet disc. Since there is no correlation between the orientations
of the bullet disc, the target and the encounter orbit, this configuration
reflects an uncorrelated or random in-flow of material.
The subsequent evolution mainly depends on the amount and infall angle of
this additional material, and also on the protoplanetary evolution
stage during the capture.

In our calculations we focus on the case of
an encounter before any planet has formed in the passing star's
disc. It is thus a purely gaseous interaction computed via smoothed
particle hydrodynamics (SPH).
In our model, the star is a sun-type star with a low-mass PPD
($\mdisk\le0.1\,\msun$, radius $<100$~AU), passing through the target
disc (0.5~\tmsun).

The initial conditions for the PPD are taken from \citet{StaWhi08,StaWhi09a}.
The disc model has power-law 
profiles for temperature, $T$, 
and surface density, $\Sigma$, from \citet{StaWhi08}, both as a
function of the distance, $R$, from the central star
\begin{equation}\label{eq:sigma}
\Sigma(R)=\Sigma_0\,\left(\frac{R}{\mathrm{AU}}\right)^{-\psig}\,,
\end{equation}
\begin{equation}\label{eq:temp}
T(R)=\left[T_0^2\,\left(\frac{R}{\mathrm{AU}}\right)^{-2\pt}+T_\infty^2\right]^{1/2}\,,
\end{equation}
where $1.5\le\psig\le1.75$ and $0.5\le\pt\le1.0$. $T_\infty=10$~K is
a background temperature to account for background radiation from
other stars within the host cluster.
The radiation temperature of the stars are estimated assuming a
mass-to-luminosity relation of $L\propto M^4$ for low-mass stars.

The disc is initially populated between 4 and 40~AU. Preceding the
actual encounter the disc is allowed to ``settle'' during the
approach time of about
5000 years (equivalent to 20 orbits at the outer rim), erasing
artefacts from the initial distribution function. During this settling the
disc smears out at the inner and the outer border extending its radial
range from a few AU up to $\gtrsim 60$~AU. The actual surface density
after the settling and thus the mass content within a given radius is
therefore slightly smaller than in the initial setup.

Similarly, the target is set up as a massive extended disc 
with mass of 0.48~\tmsun\ and a radius of $\approx400$~AU, as well as a 0.75~\tmsun\
star to keep the gas gravitationally bound. The setup is therefore
equivalent to those used in \citet{Thiesetal2010}, where triggered
fragmentation inside this envelope
upon tidal perturbations have been studied, and is indeed realistic as long as encounters with
large circumstellar discs are considered. However, future
models will also use different setups (protostellar cores, dense gas
filaments etc.) as the target.

Both systems are initially set on a hyperbolic orbit around their centre of mass
with the discs being inclined mutually as well as with respect to the
orbital plane. The periapsis is set to 500~AU initially. The eccentricity is set to 1.1,
corresponding to a pre-encounter relative velocity of about 0.7\unit{km^{-1}}.
The periapsis relative velocity is 2.9\unit{km^{-1}}.
The hyperbolic orbit is inclined 15 degrees against the target disc plane using Eulerian
$z$-$x$-$z$ rotation angles 0, 15\grad, 0, while the
plane of the bullet disc (initially in the $x$-$y$ plane) is tilted by 0, 135\grad, 60\grad.

The eccentricity value is in agreement with the likelihood estimation of encounter
parameters by \citet{OPE10}. In addition, encounters at distances of the order of
a few hundred AU are likely to happen in clusters comparable to the Orion Nebula Cluster
\citep{TKT05,OPE10}, thus our scenario is rather likely than exceptional.
The actual periapsis and eccentricity
may differ slightly due to tidal angular momentum transfer, but these effects on the
orbit are negligible at that time.
After the passage, when both stars have reached a sufficient
distance ($\sim1000$~AU) again, the gas particles around the bullet star
(i.e. the remains of its own disc and any captured gas) are separated
from the system and analysed in the centre-of-mass system of the bullet star.

\section{Numerical method}\label{sec:methods}
All computations were performed by using the well-tested smoothed-particle hydrodynamics (SPH)
code {\sc dragon} by \citet{Goetal04}
including the radiative heat transfer
extension by \citet{Staetal07}. Most of the numerical
parameter settings have been adopted from \citet{StHuWi07,StaWhi09a}.
The radiative-transfer algorithm is not
strictly required in our model, but due to its high numerical efficiency it does not significantly
slow down the calculations, so we left it in to keep it consistent with
previous calculations by \citet{Thiesetal2010}.
All gas particles have the same mass and equation of state.
The artificial
viscosity parameters are $\alpha=0.1$ and $\beta=0.2$, i.e. lower then the usually
used values (1 and 2, respectively, \citealt{monaghan1992}), to keep the disc dispersion low.
Stars and planets
are treated by sink particles which are set in the initial 
conditions, but may also form if the local volume density exceeds
$10^{-9}\unit{g\,cm^{-3}}$ (not expected in the current model, however).
The sink radius is chosen as 0.5~AU while the sink masses are the masses of the stars
involved, i.e. 1.0~\tmsun\ for the bullet star and 0.75~\tmsun\ for the host star
of the target disc.
Any gas particle that becomes gravitationally bound to a sink within less than
the sink radius is treated as being accreted by the sink.
The code currently does not treat magnetic field nor dust.

The numerical model used here is the
same as used in \citet{Thiesetal2010}, using 250~000 
SPH particles for the target while the PPD of the passing star has been
set up by 25~000 to 50~000 particles, for disc masses between 0.05 to 0.1~\tmsun,
respectively.

\section{Results}\label{sec:results}
\begin{figure}
\begin{center}
\vspace*{0.5cm}
\includegraphics[width=0.45\textwidth]{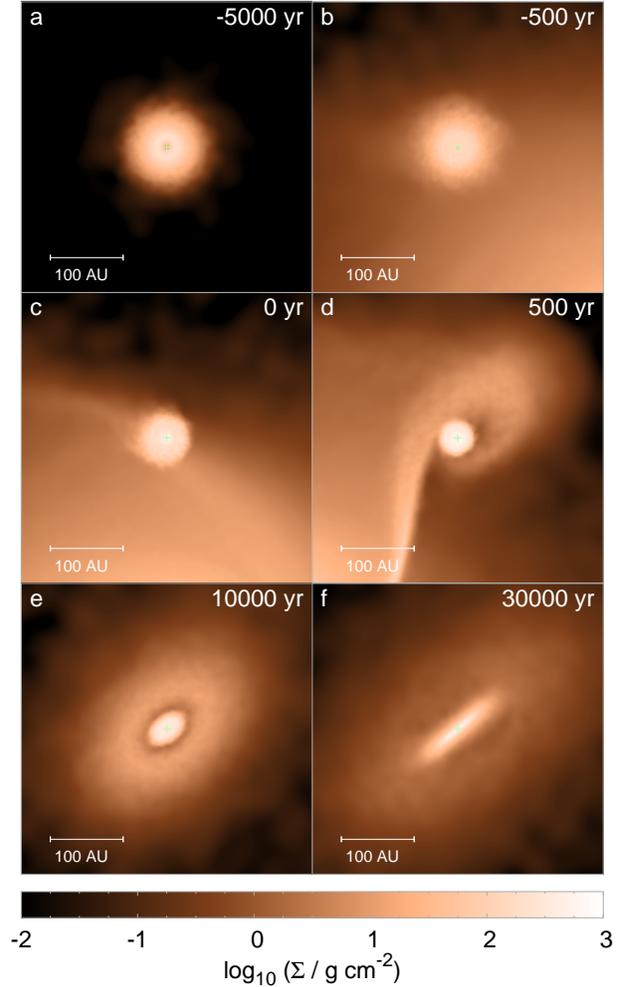}
\caption{\label{x100i}
Snapshot of a bullet star with a 0.1~\tmsun\ circumstellar disc passing through
a massive extended (0.5~\tmsun, 400~AU) target disc on an inclined orbit.
While the original disc is partially truncated, partially condensed to the
center, and somewhat tilted, the bullet star accretes additional material
from the target disc, forming an inclined annulus of gas around it (frames b to d).
The initially strongly inclined structures progressively align each other
within 30~000 years after the encounter (frames e, f).
The time stamp in each frame refers to the time of the encounter.}
\end{center}
\end{figure}

\begin{figure}
\begin{center}
\includegraphics[width=0.45\textwidth]{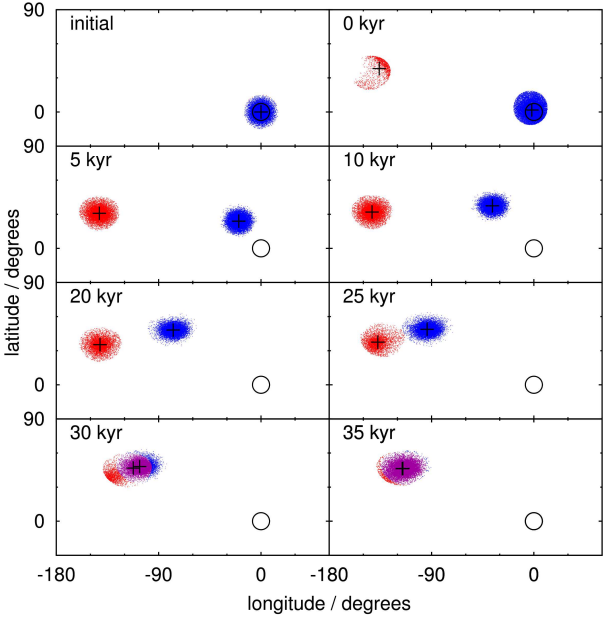}
\caption{\label{twoalign}
The angular momentum unit vectors of the particles of the material around the
bullet star in spherical coordinate angles, $\vartheta$ (latitude) and
$\varphi$ (longitude), i.e. projected on the unit sphere. The reference frame
is adjusted so that the normal vector of the bullet disc points at
$(\varphi, \vartheta)=(0, 0)$.
The topmost two frames show the angular momenta initially and at the time of
the closest approach.
While the original bullet star disc is altered only slightly there is
clearly another population of particles settling in an annulus which is
tilted against the initial disc plane. After formation, both
structures progressively align to each other, and the angular
difference between both degrades (subsequent frames).
The total angular momentum vector of each
disc is marked by a cross, while the open circle indicates the initial angular
momentum vector of the bullet star disc before the encounter (being in fact nearly
identical to that at the moment of the encounter, $t=0$). This also denotes the
direction of the bullet's star spin vector.
The time stamp in each frame refers to the time of the encounter.}
\end{center}
\end{figure}

\subsection{General effects}
\label{ssec:general}
\begin{figure}
\vspace*{0.5cm}
\begin{center}
\includegraphics[width=0.45\textwidth]{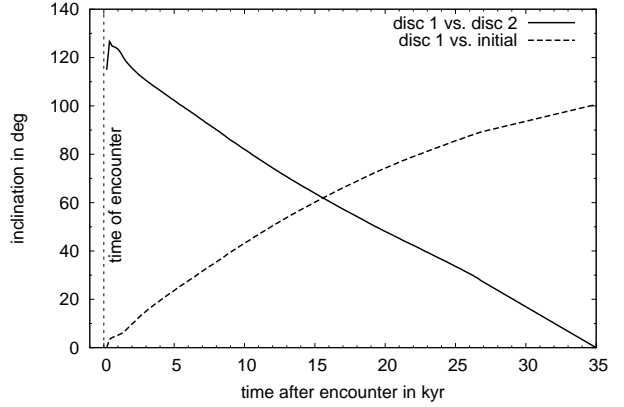}
\caption{\label{twoaligncurve}The mutual angular tilt between the pre-existing
disc of the bullet star and the annulus of captured gas,
as well as the net angular tilt of the
pre-existing disc as a function of time. While being almost
retrograde wrt. each other before the encounter,
the two structures progressively align
with each other over time. 25~kyr after the encounter the mutual tilt has
degraded to about 35 degrees. At the same time, the pre-existing disc has
been tilted wrt. its initial orientation by about 90 degrees.
It has become retrograde wrt. the bullet star's spin after about 27~000~yr.}
\end{center}
\end{figure}

Figure \ref{x100i} shows six snapshots of the encounter
and subsequent gas capture.
The view is centred on the passing star (the ``bullet star'') at
face-on direction towards the unperturbed disc.  The time stamp refers
to the moment of the periastron passage. Frame (a) shows the
unperturbed disc. Its core has a radius of about 40~AU while the
low-density outer regions extent to about 100~AU. Shortly after the
encounter, as shown in frame (b--d), the captured material starts to form
an annulus around the star while, at the same time, the original bullet disc
is partly truncated, with the inner region being condensed. The
persistent quadrupole forces continue to alter the disc's orientation
and eventually tilt it into edge-on orientation (frames e and f).
At the same time, the
orientation of the annulus is only slightly changed due to its larger
radius and thus higher moment of inertia. As clearly visible,
both the captured annulus and the disc are progressively aligning each
other, eventually forming a combined disc. The dynamical evolution of
the mutual orientation and the total tilt of the original bullet disc is
depicted in Figures \ref{twoalign} and \ref{twoaligncurve}.
Here, the SPH particles have been separated into groups sharing
a similar angular momentum unit vector and thus belonging to the
same rotational structure. Particles which deviate more than 15 degrees
from the bulk angular momentum as well as unbound or loosely bound particles
beyond 100~AU from the bullet star are neglected.
The mutual inclination of both
structures is declining from initially 130 degrees to near zero
degrees within 35~kyr. On the other hand, the original disc is tilted
by nearly 100 degrees, i.e., {\it if being initially aligned to the stellar
equator it is now retrograde.}

During the encounter, typically around 10 to 30~\tmjup\ of gas is
captured by the bullet star (Figure \ref{x100i}, frames b--d;
Section \ref{sec:accrete}).
In the cases where a protoplanetary disc (hereafter called the original disc)
is already present, two general effects can be distinguished: (i) During the
capture process, the original disc is compressed to about one-half or one-third
of the initial radius, accordingly increasing the surface density.
(ii) The captured material settles in an outer annulus the inclination
of which depends on the orbital orientation and the orientation
of the target disc relative to the encounter orbit.

The orientation of this annulus, shown in Figure \ref{twoalign} in spherical
coordinates, is not constant, but changes over time.
Within about a few thousand years after formation the annulus and the bullet disc
have considerably lowered their mutual inclination (third and fourth frame),
as shown in Figure~\ref{twoaligncurve}.
A few ten thousand years after the encounter,
they have almost aligned to each other,
eventually forming a single circumstellar disc.
Consequently, the orientation of this disc
differs considerably from its initial orientation and therefore from the
stellar equatorial plane.
It should be noted that this tilt may be much lower if a steady and homogeneous
gas flow is assumed \citep{MoeThr09}.

\subsection{Captured mass}\label{sec:accrete}
\begin{figure}
\begin{center}
\includegraphics[width=0.45\textwidth]{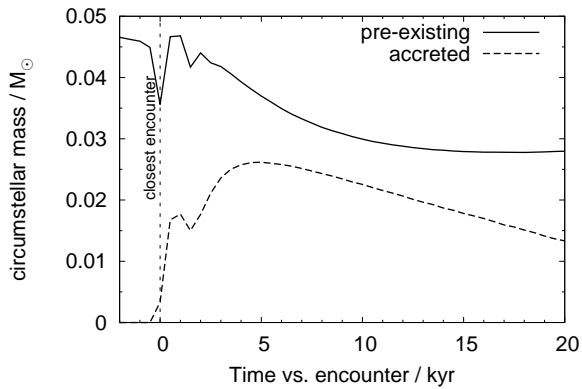}
\caption{\label{accrete}The mass of the pre-existing bullet star disc
(solid curve) and of the captured annulus (dashed curve;
see Section \ref{ssec:general}) as a function of time.
The peaks in the solid curve at about the moment of the closest encounter
are due to the perturbations of the pre-existing disc caused by the
impact of in-flowing gas and tidal forces by the target mass.
The capture process starts shortly before
the encounter and continues to drag material from the tidal arm
(Figure \ref{x100i}) even several kyr later. The slight decrease of mass
of both structures is mainly caused by dynamical scattering and, in particular
for the inner disc, accretion onto the star, but also by a small amount
due the numerical diffusion that is unavoidable in SPH calculations.
In a real disc the final masses would thus be somewhat larger.}
\end{center}
\end{figure}

The amount of captured mass is an important quantity to judge its
impact on planet formation.
In our calculations, typically between 0.01 and 0.02~\tmsun\ is captured from the
target disc, i.e. about one third of the initial bullet star disc mass.
Figure \ref{accrete} depicts the time evolution of the masses of the
bullet star disc and the captured annulus. 
As shown here, the capture of material from the target
disc begins quickly about the closest encounter, and is almost finished
within about 3000~yr. At the same time, the pre-existing disc suffers
a radial compression (see frames b--d in Figure \ref{x100i}).
The distinction of these two structures has been done by the angular momentum
grouping method described in Section \ref{ssec:general}.
It has to be noted that the curves in Figure \ref{accrete} show
a representative fraction of each gas body defined by this method.
Since part of the gas particles are also subject to scattering into large
separations ($>100$~AU) and/or inclinations ($>15$~degrees)
these are omitted by this selection process, thus
leading to a decreasing mass of each disc.
These losses are in part due
to dynamical scattering and frictional heating of the gas, in part due to
accretion by the bullet star. To a certain degree, they are also due to
numerical diffusion, which is unavoidable in SPH, especially for relatively
low resolution, and may thus be reduced by using larger particle numbers
in future calculations. However, the mass-loss does not affect the central
conclusion of this paper in any way.

\subsection{Pre-existing planets}\label{sec:planets}
\begin{table}
\begin{center}
\begin{tabular}{ccccccc}\hline
      &\multicolumn{3}{c}{initial}&\multicolumn{3}{c}{20~kyr after flyby}\\
Object&$\frac{m_{\text{ini}}}{\mjup}$&$\frac{a_{\text{ini}}}{\text{[AU]}}$&$\ecc_{\text{ini}}$ &$\frac{m_{\text{fin}}}{\mjup}$ &$\frac{a_{\text{fin}}}{\text{[AU]}}$&$\ecc_{\text{fin}}$\\\hline
1     &1.0   &5.2  &0.05 &2.8 &1.6 &0.67\\
2     &0.30  &9.5  &0.05 &3.1 &5.1 &0.20\\
3     &0.046 &19.2 &0.05 &3.2 &6.5 &0.70\\
4     &0.054 &30.1 &0.01 &2.2 &2.2 &0.89\\\hline
\end{tabular}
\caption{\label{x200aei}Masses and orbital elements of four model planets
before and after a flyby and capture event equivalent to the scenario in
Fig. \ref{x100i}. The planets, which replace the bullet star disc
of the first scenario, are initially equivalent to Jupiter, Saturn, Uranus and
Neptune. 20~kyr after the flyby, when the capture and accretion process is essentially
over, the orbits are shrunk to a fraction of their initial values while being
highly eccentric now. Furthermore, the orbital positions of planets 2--4 have
been altered. The system has apparently become unstable.}
\end{center}
\end{table}
We also studied the effects of
such a gas capture event on a pre-existing planetary system.
Here, the circumstellar disc of the bullet star has been
replaced by a model planetary system equivalent Jupiter, Saturn, Uranus,
and Neptune. Like the stars, the planets are represented by sink particles
with a sink radius of 0.5~AU. This value is a compromise between spatial
resolution and mass resolution. Gas particles passing through the sphere
defined by the sink radius
are considered as captured if they are gravitationally bound to the sink,
otherwise they simply pass through it.
The current results, summarised in Table \ref{x200aei},
show that a planetary system equivalent to the Sun orbited by Jupiter,
Saturn, Uranus, and Neptune (with initial orbital radii of 5, 10, 20, and
30~AU, respectively) is severely affected by gas-planet
interaction. During the capture some 0.01~\tmsun\ of gas flow towards
the bullet star, crossing the planetary orbits. In response, the planets
migrate inward to semimajor axes of about one tenth their initial
separations, i.e. between about 0.5 and 4~AU. In addition, some orbits
get strongly eccentric (up to $e=0.8$, where $e=0$ means a circular
orbit and $e\ge1.0$ means ejection). Such a system is
dynamically unstable, i.e. some planets are probably ejected over
time, typically leaving one to three planets, with the most massive
planet on a close-in orbit. Systems like these are expected to undergo
significant subsequent evolution. Even after ejection of the most unstable
planets the remaining ones may still be subject to Kozai resonance,
altering their eccentricities and inclinations over many orbital times.
But it is clear that these results suggest that misaligned planets and
hot Jupiters as well as short-period planets on eccentric orbits may
both be an outcome of gas flow onto a forming or existing system.

\subsection{Stellar spin}
The current work assumes that the stellar spin is not largely changed
since the assembly of its circumstellar disc, and thus its angular momentum
vector points into approximately the same direction as the angular momentum
vector of the disc before the encounter. This assumption, however, may not
be true in general. As \citet{Laietal11} have found, the stellar spin may,
in some cases, be tilted through magnetic interaction with the disc.
Furthermore, during the early phases, when the protostar has still a considerable
diameter (tenths of an AU), it may still re-align to the disc after an
encounter event of the kind proposed in our paper. However, a large fraction
of stars will already be compact enough to be effectively decoupled from
angular momentum transfer with their disc, given the absence of strong magnetic
torque. Since encounter events and gas capture are a natural consequence of
star formation in dense environments, we conclude that they indeed do play an
important role in forming mis-aligned planetary systems.

\subsection{Resolution issues}\label{ssec:resolution}
The required resolution strongly depends on the kind of scenario to be
modelled. For example, for studying fragmentation the local Jeans mass
has to be sufficiently resolved, meaning by at least about $~50$
SPH particles. The study of the global behaviour of gas masses without
focus on fragmentation, on the other hand, requires much less
resolution. In our model, the mass of each gas particle is
$1.92\dez{-6}\,\msun$ or 0.64~$\mearth$. The well-resolved mass
therefore is about 32~$\mearth$ or twice the mass of Neptune. The
transferred mass in both models (encounter of disc with target and
planetary-system vs. target, respectively) is of the order of 0.01~\tmsun\
or 5000 SPH particles, so mass resolution is not an issue here.
For calculating the impact on dust components
or to study vortex formation \citep{BaSo1995,KlahrBodenheimer2004} this
resolution is likely to be insufficient. These issues will be addressed in
future work.

One could argue that the masses of Uranus and Neptune equivalents
(planets 3 and 4, see Section \ref{sec:planets}),
which are roughly equal to the interacting mass required
for efficient momentum transfer, are less-optimally
resolved by only 20 to 25 SPH particles. The statistical scatter in terms
of the number of gas particles interacting with these minor giant planets
is about $\sqrt{1/25}=20\pct$ and therefore does not significantly
influence the outcome of drag-induced migration. Furthermore, this behaviour
is fully consistent with the migration of the Jupiter and Saturn equivalents
(planets 1 and 2). A word of caution has, however, to be stated at this point:
Apart from mere resolution issues the proper treatment of accretion onto
sinks shows a number of pitfalls. For example, the void of material in the
sink may lead to artificial suction effects and thus speed-up accretion.
The time-evolution of the planetary orbits has therefore to be treated with
great caution. The long-term outcome, however, largely depends on basic physical
principles like conservation of angular momentum and can therefore be
expected to reflect the real consequences on a pre-existing planetary system.

\section{Discussion and conclusions}\label{sec:conclusions}
We have analysed the general effects of inclined gas capture onto a
pre-existing circumstellar disc upon passage through a dense gaseous
reservoir. The secondary gas capture does
not destroy a pre-existing disc even if the new material
inflows at a highly inclined angle. Rather, the inner regions of the disc
(inside about 30 AU) become denser, while some of the outer disc
material is scattered away.
The major result is the formation of an inclined combined
circumstellar disc from captured material and the pre-existing disc.
After capturing a few 0.01~\tmsun\ into an initially
inclined annulus both structures tend to align with each other within a
few 10~000 years after the encounter. An annulus with an initial
inclination of $135\grad$ aligns to about $35\grad$ within 20~000 years.
Due to conservation of angular momentum this results in a net shift
of more than $90\grad$ in our model, and even more for lower-mass
pre-existing discs or disc-less stars. If planets form from the resulting
inclined disc one naturally arrives at misaligned planets.

The resulting disc may have two chemically distinct radial regions, the inner
region which is associated with the star, and the outer region composed from
the captured material.

In addition, the basic effect of capture onto a pre-existing planetary
system have been estimated. Orbits typically shrink and become strongly
eccentric in response to the contact with captured gas and subsequent
planet-planet interaction,
while their orbital planes can be dramatically altered thus naturally leading
to (even mutually) misaligned short-period planets with eccentric orbits.

Important constraints are given by recent observations of apparent
inclinations between circumstellar discs and their host stars.
In a most recent analysis \citet{Watsonetal2011} have found no significant
misalignment between the normal vectors of debris discs and the spin axis
of low-mass to solar-type stars. In particular, the projected inclinations
between the star and its disc typically differ by 5--10 degrees,
and 15--45 degrees in a few cases.
Their results do not necessarily contradict our scenario since even moderate
disc tilts and warps may lead to a Kozai resonance in a planetary system
born out of such a disc. Following the mechanism described by \citet{FabTre07}
and \citet{Naozetal2011} near-random misalignment of a close-in planet and the
spin of its host star may result.

Future studies will also consider protoplanetary discs with embedded
planets and systems with a brown dwarf on a wide orbit (about 100--200~AU).
In addition, the consequences for dust content and mixing of
differently composed material will be analysed.

The mechanism described in this paper provides a natural scenario for
the formation of misaligned planetary systems in gas-rich dense
star-forming regions. While other models have been
proposed our scenario requires rather simple assumptions, and may even provide
the initially misaligned planetary orbits required by the recent model of
\citet{Naozetal2011}.

\section*{Acknowledgements}
This project has been funded by DFG grant KR1635/25 as part of the SPP1385.


\begin{thebibliography}{}

\bibitem[\protect\citeauthoryear{{Anderson}, {Hellier}, {Gillon}, {Triaud},
  {Smalley}, {Hebb}, {Collier Cameron}, {Maxted} et~al.,}{{Anderson}
  et~al.}{2010}]{wasp17b_01}
{Anderson} D.~R.,  {Hellier} C.,  {Gillon} M.,  {Triaud} A.~H.~M.~J.,
  {Smalley} B.,  {Hebb} L.,  {Collier Cameron} A.,  {Maxted}   et~al., 2010,
  \apj, 709, 159

\bibitem[\protect\citeauthoryear{{Barge} \& {Sommeria}}{{Barge} \&
  {Sommeria}}{1995}]{BaSo1995}
{Barge} P.,  {Sommeria} J.,  1995, \aap, 295, L1

\bibitem[\protect\citeauthoryear{{Bate}, {Bonnell} \& {Bromm}}{{Bate}
  et~al.}{2003}]{BBB03}
{Bate} M.~R.,  {Bonnell} I.~A.,    {Bromm} V.,  2003, \mnras, 339, 577

\bibitem[\protect\citeauthoryear{{Bate}, {Lodato} \& {Pringle}}{{Bate}
  et~al.}{2010}]{BLP10}
{Bate} M.~R.,  {Lodato} G.,    {Pringle} J.~E.,  2010, \mnras, 401, 1505

\bibitem[\protect\citeauthoryear{{Boss}}{{Boss}}{2004}]{2004ApJ...610..456B}
{Boss} A.~P.,  2004, \apj, 610, 456

\bibitem[\protect\citeauthoryear{{Fabrycky} \& {Tremaine}}{{Fabrycky} \&
  {Tremaine}}{2007}]{FabTre07}
{Fabrycky} D.,  {Tremaine} S.,  2007, \apj, 669, 1298

\bibitem[\protect\citeauthoryear{{Gillon}}{{Gillon}}{2009}]{Gillon09}
{Gillon} M.,  2009, MNRAS, preprint (arXiv:0906.4904)

\bibitem[\protect\citeauthoryear{{Goodwin}, {Whitworth} \&
  {Ward-Thompson}}{{Goodwin} et~al.}{2004}]{Goetal04}
{Goodwin} S.~P.,  {Whitworth} A.~P.,    {Ward-Thompson} D.,  2004, \aap, 414,
  633

\bibitem[\protect\citeauthoryear{{H{\'e}brard}, {Bouchy}, {Pont}, {Loeillet},
  {Rabus}, {Bonfils}, {Moutou}, {Boisse} et~al.,}{{H{\'e}brard}
  et~al.}{2008}]{Heetal08}
{H{\'e}brard} G.,  {Bouchy} F.,  {Pont} F.,  {Loeillet} B.,  {Rabus} M.,
  {Bonfils} X.,  {Moutou} C.,  {Boisse}   et~al., 2008, \aap, 488, 763

\bibitem[\protect\citeauthoryear{{Johnson}, {Winn}, {Albrecht}, {Howard},
  {Marcy} \& {Gazak}}{{Johnson} et~al.}{2009}]{Joetal09}
{Johnson} J.~A.,  {Winn} J.~N.,  {Albrecht} S.,  {Howard} A.~W.,  {Marcy}
  G.~W.,    {Gazak} J.~Z.,  2009, \pasp, 121, 1104

\bibitem[\protect\citeauthoryear{{Klahr} \& {Bodenheimer}}{{Klahr} \&
  {Bodenheimer}}{2004}]{KlahrBodenheimer2004}
{Klahr} H.,  {Bodenheimer} P.,  2004, in {Garca-Segura} G.,  {Tenorio-Tagle}
  G.,  {Franco} J.,   {Yorke} H.,  eds, Revista Mexicana de Astronomia y
  Astrofisica Conference Ser., {Tornados and Hurricanes in Planet Formation}.
Instituto de Astronoma, UNAM, Mexico, pp 87--90

\bibitem[\protect\citeauthoryear{{Lai}, {Foucart} \& {Lin}}{{Lai}
  et~al.}{2011}]{Laietal11}
{Lai} D.,  {Foucart} F.,    {Lin} D.~N.~C.,  2011, \mnras, 412, 2790

\bibitem[\protect\citeauthoryear{{McArthur}, {Benedict}, {Barnes}, {Martioli},
  {Korzennik}, {Nelan} \& {Butler}}{{McArthur}
  et~al.}{2010}]{2010ApJ...715.1203M}
{McArthur} B.~E.,  {Benedict} G.~F.,  {Barnes} R.,  {Martioli} E.,  {Korzennik}
  S.,  {Nelan} E.,    {Butler} R.~P.,  2010, \apj, 715, 1203

\bibitem[\protect\citeauthoryear{{McLaughlin}}{{McLaughlin}}{1924}]{McLaughlin%
1924}
{McLaughlin} D.~B.,  1924, \apj, 60, 22

\bibitem[\protect\citeauthoryear{{Moeckel} \& {Throop}}{{Moeckel} \&
  {Throop}}{2009}]{MoeThr09}
{Moeckel} N.,  {Throop} H.~B.,  2009, \apj, 707, 268

\bibitem[\protect\citeauthoryear{{Monaghan}}{{Monaghan}}{1992}]{monaghan1992}
{Monaghan} J.~J.,  1992, \araa, 30, 543

\bibitem[\protect\citeauthoryear{{Mordasini}, {Alibert}, {Benz} \&
  {Naef}}{{Mordasini} et~al.}{2008}]{2008ASPC..398..235M}
{Mordasini} C.,  {Alibert} Y.,  {Benz} W.,    {Naef} D.,  2008, in {D.~Fischer,
  F.~A.~Rasio, S.~E.~Thorsett, \& A.~Wolszczan} ed., Astronomical Society of
  the Pacific Conference Series Vol.~398 of Astronomical Society of the Pacific
  Conference Series, {Giant Planet Formation by Core Accretion}.
pp 235--+

\bibitem[\protect\citeauthoryear{{Naoz}, {Farr}, {Lithwick}, {Rasio} \&
  {Teyssandier}}{{Naoz} et~al.}{2011}]{Naozetal2011}
{Naoz} S.,  {Farr} W.~M.,  {Lithwick} Y.,  {Rasio} F.~A.,    {Teyssandier} J.,
  2011, Nature, 473, 187

\bibitem[\protect\citeauthoryear{{Narita}, {Sato}, {Hirano} \&
  {Tamura}}{{Narita} et~al.}{2009}]{Naritaetal09}
{Narita} N.,  {Sato} B.,  {Hirano} T.,    {Tamura} M.,  2009, \pasj, 61, L35

\bibitem[\protect\citeauthoryear{{Olczak}, {Pfalzner} \& {Eckart}}{{Olczak}
  et~al.}{2010}]{OPE10}
{Olczak} C.,  {Pfalzner} S.,    {Eckart} A.,  2010, \aap, 509, A63+

\bibitem[\protect\citeauthoryear{{Pfalzner}, {Tackenberg} \&
  {Steinhausen}}{{Pfalzner} et~al.}{2008}]{PTS08}
{Pfalzner} S.,  {Tackenberg} J.,    {Steinhausen} M.,  2008, \aap, 487, L45

\bibitem[\protect\citeauthoryear{{Pont}, {Endl}, {Cochran}, {Barnes}, {Sneden},
  {MacQueen}, {Moutou}, {Aigrain} et~al.,}{{Pont} et~al.}{2010}]{Pontetal10}
{Pont} F.,  {Endl} M.,  {Cochran} W.~D.,  {Barnes} S.~I.,  {Sneden} C.,
  {MacQueen} P.~J.,  {Moutou} C.,  {Aigrain}   et~al., 2010, \mnras, 402, L1

\bibitem[\protect\citeauthoryear{{Pont}, {H{\'e}brard}, {Irwin}, {Bouchy},
  {Moutou}, {Ehrenreich}, {Guillot}, {Aigrain} et~al.,}{{Pont}
  et~al.}{2009}]{Pontetal09}
{Pont} F.,  {H{\'e}brard} G.,  {Irwin} J.~M.,  {Bouchy} F.,  {Moutou} C.,
  {Ehrenreich} D.,  {Guillot} T.,  {Aigrain}   et~al., 2009, \aap, 502, 695

\bibitem[\protect\citeauthoryear{{Rossiter}}{{Rossiter}}{1924}]{Rossiter1924}
{Rossiter} R.~A.,  1924, \apj, 60, 15

\bibitem[\protect\citeauthoryear{{Stamatellos}, {Hubber} \&
  {Whitworth}}{{Stamatellos} et~al.}{2007a}]{StHuWi07}
{Stamatellos} D.,  {Hubber} D.~A.,    {Whitworth} A.~P.,  2007a, \mnras, 382,
  L30

\bibitem[\protect\citeauthoryear{{Stamatellos}, {Whitworth}, {Bisbas} \&
  {Goodwin}}{{Stamatellos} et~al.}{2007b}]{Staetal07}
{Stamatellos} D.,  {Whitworth} A.~P.,  {Bisbas} T.,    {Goodwin} S.,  2007b,
  \aap, 475, 37

\bibitem[\protect\citeauthoryear{{Stamatellos} \& {Whitworth}}{{Stamatellos} \&
  {Whitworth}}{2008}]{StaWhi08}
{Stamatellos} D.,  {Whitworth} A.~P.,  2008, \aap, 480, 879

\bibitem[\protect\citeauthoryear{{Stamatellos} \& {Whitworth}}{{Stamatellos} \&
  {Whitworth}}{2009}]{StaWhi09a}
{Stamatellos} D.,  {Whitworth} A.~P.,  2009, \mnras, 392, 413

\bibitem[\protect\citeauthoryear{{Thies} \& {Kroupa}}{{Thies} \&
  {Kroupa}}{2007}]{TK07}
{Thies} I.,  {Kroupa} P.,  2007, \apj, 671, 767

\bibitem[\protect\citeauthoryear{{Thies}, {Kroupa}, {Goodwin}, {Stamatellos} \&
  {Whitworth}}{{Thies} et~al.}{2010}]{Thiesetal2010}
{Thies} I.,  {Kroupa} P.,  {Goodwin} S.~P.,  {Stamatellos} D.,    {Whitworth}
  A.~P.,  2010, \apj, 717, 577

\bibitem[\protect\citeauthoryear{{Thies}, {Kroupa} \& {Theis}}{{Thies}
  et~al.}{2005}]{TKT05}
{Thies} I.,  {Kroupa} P.,    {Theis} C.,  2005, \mnras, 364, 961

\bibitem[\protect\citeauthoryear{{Triaud}, {Collier Cameron}, {Queloz},
  {Anderson}, {Gillon}, {Hebb}, {Hellier}, {Loeillet}, {Maxted}, {Mayor},
  {Pepe}, {Pollacco}, {S{\'e}gransan}, {Smalley}, {Udry}, {West} \&
  {Wheatley}}{{Triaud} et~al.}{2010}]{Triaudetal2010}
{Triaud} A.~H.~M.~J. et al.,  2010, \aap, 524, A25

\bibitem[\protect\citeauthoryear{{Watson}, {Littlefair}, {Diamond}, {Collier
  Cameron}, {Fitzsimmons}, {Simpson}, {Moulds} \& {Pollacco}}{{Watson}
  et~al.}{2011}]{Watsonetal2011}
{Watson} C.~A.,  {Littlefair} S.~P.,  {Diamond} C.,  {Collier Cameron} A.,
  {Fitzsimmons} A.,  {Simpson} E.,  {Moulds} V.,    {Pollacco} D.,  2011,
  \mnras, 413, L71

\bibitem[\protect\citeauthoryear{{Winn}, {Johnson}, {Fabrycky}, {Howard},
  {Marcy}, {Narita}, {Crossfield}, {Suto}, {Turner}, {Esquerdo} \&
  {Holman}}{{Winn} et~al.}{2009a}]{Winetal09a}
{Winn} J.~N.,  {Johnson} J.~A.,  {Fabrycky} D.,  {Howard} A.~W.,  {Marcy}
  G.~W.,  {Narita} N.,  {Crossfield} I.~J.,  {Suto} Y.,  {Turner} E.~L.,
  {Esquerdo} G.,    {Holman} M.~J.,  2009a, \apj, 700, 302

\bibitem[\protect\citeauthoryear{{Winn}, {Johnson}, {Albrecht}, {Howard},
  {Marcy}, {Crossfield} \& {Holman}}{{Winn} et~al.}{2009b}]{Winetal09b}
{Winn} J.~N.,  {Johnson} J.~A.,  {Albrecht} S.,  {Howard} A.~W.,  {Marcy}
  G.~W.,  {Crossfield} I.~J.,    {Holman} M.~J.,  2009b, \apjl, 703, L99

\end{thebibliography}
\end{document}